\documentclass[aps,prd,twocolumn,superscriptaddress,nofootinbib,longbibliography]{revtex4-1}
\usepackage{amsmath,amssymb, amsthm,amstext}
\usepackage{graphicx}
\usepackage{color}
\usepackage{array, enumerate}
\usepackage{bm}
\usepackage[breaklinks,colorlinks,citecolor=blue]{hyperref}
\usepackage{braket}
\usepackage{txfonts}
\usepackage[normalem]{ulem}

\newcommand{\dd}[0]{\mathrm{d}}

\def\be{\begin{equation}}
\def\ee{\end{equation}}

\def\apjl{Astroph.J.Lett.}

\def\mnras{MNRAS}

\def\aap{Astronomy\&Astrophysics}

\begin{document}

%\title{Secondary spins of extreme mass-ratio inspirals: \\ 
%Measurement precision and astrophysical applications}
\title{Secondary spins of extreme mass-ratio inspirals: \\ 
A probe to the formation channels}
\author{Qiuxin Cui}
%\email{cuiqiuxin@shao.ac.cn}
\affiliation{Shanghai Astronomical Observatory, 80 Nandan Road, Shanghai, 200030, People’s Republic of China}
\author{Wen-Biao Han }
\email[Contact author:]{wbhan@shao.ac.cn}
\affiliation{Shanghai Astronomical Observatory, 80 Nandan Road, Shanghai, 200030, People’s Republic of China}
\author{Zhen Pan}
\email[Contact author:]{zhpan@sjtu.edu.cn}
\affiliation{Tsung-Dao Lee Institute, Shanghai Jiao-Tong University, Shanghai, 520 Shengrong Road, 201210, People’s Republic of China}
\affiliation{School of Physics \& Astronomy, Shanghai Jiao-Tong University, Shanghai, 800 Dongchuan Road, 200240, People’s Republic of China}
\begin{abstract}
 Extreme mass-ratio inspirals (EMRIs), consisting of a secondary (stellar mass) black hole (BH) orbiting around a supermassive BH, are one of the primary targets for future spaceborne gravitational wave (GW) detectors. The spin of the secondary BH encodes the formation history of the stellar mass BH
 and the formation process of the EMRI.
 In this work, we construct a kludge EMRI waveform model taking the secondary spin into account and \textcolor{red}{preliminarily} forecast the measurement precision of the  secondary spin by future spaceborne GW detectors with the Fisher information matrix. \textcolor{red}{We find the secondary spin might be measured with reasonably good precision for generic eccentric and inclined EMRIs, with the caveat that the predictive precision may be constrained by the model's inherent simplifications.}
 As an example of its astrophysical applications, we propose that the secondary spin can be used for distinguishing dry (loss cone)  EMRIs (where the secondary 
 BHs were born in the collapse of individual massive stars and are of low spin)
 and Hills EMRIs (where the secondary BHs are remnants of massive star binaries and the secondary spins follow a bimodal
distribution). 
\end{abstract}
\date{\today}

\maketitle

\section{Introduction}

Since the first direct detection of a gravitational wave (GW) signal from a binary black hole merger in 2015 \cite{GW150914}, LIGO-Virgo-KAGRA (LVK) collaboration has announced more than 100 compact binary merger events \cite{GWTC-3}.
%, and the era of gravitational wave astronomy has officially dawned. 
The gravitational waves from these stellar-mass compact binaries at coalescence are in relative high frequency band. 
The future space-borne gravitational wave detectors, such as LISA \cite{LISA}, Taiji \cite{Taiji}, and Tian-Qin \cite{TianQin}, are 
most sensitive in the milli-Hertz (mHz) band.
%designed to observe gravitational waves in the milli-Hertz band and to study supermassive black holes, the high-redshift universe, and gravitational waves in the early universe.  
One of the primary sources for these space-borne interferometers is extreme mass-ratio inspirals (EMRI),
which consists of a stellar-mass compact object, e.g., a stellar-mass black hole (sBH) with mass $\mathcal{O}(1-100)$ $M_\odot$, and a supermassive black hole (SMBH).

The EMRI GW signals in the sensitivity band of spaceborne GW detectors  in general last for a few months to a few years depending on 
the masses of the system, and the compact object completes $\sim 10^4-10^5$ cycles around the SMBH. As a result, the intrinsic EMRI parameters (e.g., the mass and spin of SMBH, the mass of the secondary object) can be measured to a superb precision, with relative uncertainty as low as $\sim 10^{-4}-10^{-6}$ \cite{LISA_EMRI}, making EMRI a unique tool to resolve the nature of the SMBH and test general relativity \cite{TestGR_LISA}. EMRIs are  also well known as a sensitive probe to the SMBH environment \cite{Conor2025}, including 
a nearby third body \cite{Bonga:2019ycj},  the accretion disk in the case of accreting SMBH \cite{Kocsis2011,Yunes2011,Speri2023,Chat2023,Duque:2024mfw} and a possible axion-like cloud 
\cite{Zhang:2019eid,Zhang:2018kib,Brito:2023pyl,Duque:2023seg} . In this paper, we mainly focus on the spin of the secondary body, the measurement precision and the astrophysical applications,
which are largely unexplored.

Generally, the dynamic of a spinning particle in a curved spacetime is described by so-called Mathison-Papapetrous-Dixon (MPD) equations, which were proposed and developed by Mathisson \cite{mathisson1937neue} and Papapetrou \cite{Papapetrou1951} in order to solve the extended body problem in general relativity, and then reformulated by Dixon \cite{Dixon1970,Dixon1974}. As for the EMRI system, the effect of the secondary spin emerges at the first order of mass ratio, in the form of spin-curvature force following the geodesic equation. Some special orbits have been carefully studied, such as the straight falling one in Ref.~\cite{Mino1996}, and the circular-aligned one in Ref.~\cite{Tanaka1996,Han2010,Gair2011,Gair2012,Piovano2021,Jiang2024I,JiangYe2024II,Rahman2023}. So far, there is no well-established waveform template to describe the generic case, but a lot of efforts have been made recently. Ref.~\cite{Drummond2022} computed the bound orbit by associating it with a reference geodesic that shares the same turning points. Ref.~\cite{Viktor2023} calculated the asymptotic GW energy and angular momentum fluxes for generic orbits in the linear spin approximation, and Ref.~\cite{Piovano2024} improved it by using the Hamilton-Jacobi formalism found in Ref.~\cite{Witzany2019}. Ref.~\cite{Viktor2024II} calculated the post-Newtonian (PN) expansion of eccentric orbits of spinning bodies around Schwarzschild black holes and then computed the PN fluxes from these orbits. Ref.~\cite{Drummond2024} developed a model that treats the generic inspiral as a sequence of geodesic orbits. And recently, Ref.~\cite{Viktor2024} introduced a coordinate shift of a generic orbit to the virtual geodesic worldline in the linear-spin approximation, which leads to the separable formula and thus should be useful for fully relativistic waveform modeling. 

In Ref.~\cite{Gair2011}, the authors developed a kludge waveform model, which uses the MPD-based equations of motion derived by Ref.~\cite{Saijo1998} in an equatorial-aligned case for spinning black hole binaries and includes first-order conservative self-force corrections to compute the evolution of the inspiraling sBH's orbital frequency. Their analysis demonstrates that LISA is unable to measure the spin of sBH inspiraling into the Kerr SMBH with a typically mass-ratio $\sim 10^{-5}$, as the mean error is much larger than 1. Some follow-up studies confirm this conclusion, such as Ref.~\cite{Gair2012} which improves the waveform model by extending to higher-order spin effects, Refs.~\cite{Piovano2021} which uses fully relativistic numerical waveforms to the leading order in an adiabatic expansion, and Ref.~\cite{Burke2024} which employs Bayesian inference to assess the importance of first post-adiabatic terms.  However, in these researches mentioned above, the motion of the sBH is strictly constrained to an equatorial and further circular orbit, where the secondary spin must be aligned to the primary one. This constrained case may not represent the full picture of the secondary spin in real EMRI dynamics.
As we will show later, the secondary spin plays a much more important role in EMRI dynamics for generic  inclined and eccentric orbits,
therefore will be more tightly constrained from EMRI waveforms.

Actually, the secondary spin in the EMRI system contains rich astrophysical information about the sBH
formation and the EMRI formation processes. As was studied with LVK binary BH mergers in Ref.~\cite{EvolutionRoad2020}, the spin of sBH combined with the mass can reveal the evolutionary road (such as angular momentum transport) of massive stars. Beside, Ref.~\cite{Piovano2020,Lujia2025} suggested that measurements of the secondary spin may be used to distinguish the sBH from other compact objects. Similar to LVK binary BH mergers, a number of EMRI formation channels have been proposed.
The most considered EMRI sources for LISA are the dry (loss-cone) EMRIs \cite{Hopman2005,Preto2010,Bar-Or2016,Babak2017,Amaro2018,Broggi2022}, in which a stellar compact object is captured by SMBH by being pushed into a gravitational radiation-efficient orbit via multi-body scatterings within the galactic nuclei. In addition, Hills EMRIs proposed by Ref.~\cite{Hills1988,Miller2005}, in which one of the binary BHs is captured due to tidal disruption near the SMBH, should also contribute a fraction of detectable EMRIs. Another channel that contributes comes from the SMBH associated with an accretion disk, as known in active galactic nuclei (AGN), the motion of sBH is dampened in the interaction with the disk, finally captured onto the disk plane and migrates inward in the accretion disk, and thus is named wet EMRIs \cite{Sigl2007,Levin2007,Pan2021prd,Pan2021b,Pan2021,Pan2022,Derdzinski2023}.

A key question that remains to be answered is: What is the unique signature of each formation channel that is observable in future EMRI detections? 
A clear signature of wet EMRIs is the low orbital eccentricity, which is distinct from EMRIs in the other two channels. 
Hills EMRIs were previously thought to be circular in the LISA sensitivity band \cite{Miller2005}, 
therefore are obviously different from dry EMRIs.
But more recent detailed simulations show that dry and Hills EMRIs in the mHz band
are actually quite similar in their orbital properties \cite{Raveh2021}: high orbital eccentricities and no preferred orbital orientation. 
As a piece of indirect evidence for different EMRI formation channels, recent analyses of quasi-periodic eruptions, which are likely EM counterparts to low-frequency EMRIs with orbital frequency $f_{\rm obt}\sim 10^{-5}$ Hz, 
show that there are indeed clues to two different populations of EMRIs \cite{Zhou2024a,Zhou2024b,Zhou2024c}: 
a low-eccentricity population that is consistent with the wet channel prediction
and a high-eccentricity population that is consistent with both the dry and the Hills channels.
It is still unclear how to distinguish dry EMRIs from Hills EMRIs due to their similarity in orbital properties. 
In this work, we point out that 
the secondary spin can be used as a discriminator: the dry EMRIs are of low secondary spins 
due to the efficient angular momentum transport from the core to the envelope in the late evolution stage of massive stars as informed by LVK BH spins  \cite{LIGOScientific:2018jsj,Roulet:2018jbe,Fuller2019,KAGRA:2021duu},
while the Hills EMRIs are expected to be double peaked in secondary spins, because the second born BH in a BH binary is expected to spin faster due to the
tidal spin up in the Helium core stage by the first born BH \cite{Zaldarriaga:2017qkw,Bavera:2020inc,Mandel:2020lhv,Adamcewicz:2023szp}.

In this paper, we construct a kludge EMRI waveform model taking the secondary spin into account by numerically integrating MPD equations and PN radiation reactions. With this waveform model, we then forecast the  measurement precision of secondary spins by Taiji. We find the secondary spin should be measured to a reasonable precision for generic inclined and eccentric orbits while almost unconstrained for equatorial and circular orbits.
As an astrophysical application, we propose that the secondary spins can be used for distinguishing dry and Hills EMRIs:
dry EMRIs are expected to be of low secondary spins, and 
a robust detection of high secondary spins will be a clear signature of Hills EMRIs. 
Note that Taiji and LISA are pretty the same in the designed sensitivity (see Fig.~\ref{fig:wave}), so the forecast result in this work also equally applies to LISA.

\textcolor{red}{
But, the readers should pay attention to the fact that the EMRI waveform model in this paper is not self-consistent, as we exclude some same-order terms. Specifically, due to the extremely small mass ratio $\epsilon$ in EMRI systems, there exists a known two-timescale expansion \cite{LISAwaveformModel2023}. That is to say, the orbit phases $\psi_A$ evolve on the orbit time scale $\sim 2\pi/\Omega_A$ as 
\begin{equation}
\frac{d\psi_A}{dt} = \Omega^{(0)}_A\left(J^B\right) + \epsilon \Omega^{(1)}_A\left(J^B\right) + \mathcal{O}\left(\epsilon^2\right) \ ,
\end{equation}
where $\Omega_A$ are the orbit frequencies and $J^A$ are the system parameters (such as semi-latus rectum and eccentricity). While $J^A$ evolve slowly on the inspiral time scale $\sim 2\pi/(\epsilon \Omega_B)$ as 
\begin{equation}
\frac{dJ^A}{dt} = \epsilon G^A_{(1)}\left(J^B\right) + \epsilon^2 G^A_{(2)}\left(J^B\right) + \mathcal{O}\left(\epsilon^3\right) \ ,
\end{equation}
where $G^A$ are the time-averaged dissipative effects. For compact secondary objects, the spin (finite-size effect) contributes the first-order conservative and the second-order dissipative forces in $\Omega^{(1)}_A$ and $G^A_{(2)}$, which are often named the Post-1-adiabatic (1PA) terms. The 1PA terms should also include the finite-mass self-force effects, but only the conservative part of spin effects is considered in this paper. This simplified dynamics modeling may overestimate or underestimate the accuracy of the secondary spin measurements, and yet the secondary spins should still be a useful probe for formation channels in future observations. 
}

This paper is organized as follows: In Sec.~\ref{subsection:wave_model}, we start with introducing the waveform model used in this paper, and then in Sec.~\ref{subsection:precision}, we introduce the data analysis method  and we present the forecast measurement precision of the secondary spin. In Sec.~\ref{subsec:spin model}, we introduce the main feature of the three EMRI channels, then propose a distribution of the secondary spin for dry and Hills EMRIs. Finally, in Sec.~\ref{subsection:infer basic} and Sec.~\ref{subsection:inference result}, we present the detailed inference result of formation channels (the recovered distribution). We conclude this work with Sec.~\ref{section:conclusion}.
Throughout this paper, the geometrical units with $c=G=1$ and the Einstein summation convention are adopted. 

\section{Secondary spin measurement of EMRIs}

\subsection{Waveform model}\label{subsection:wave_model}
The equation of a spinning particle's motion in a curved spacetime (MPD equations) can be summarized as below, 
\begin{align}
\label{MPD1} \frac{d p^{\mu}}{d \tau} &= -\frac{1}{2} S^{\kappa \lambda} v^{\rho} R^{\mu}_{\ \rho \kappa \lambda} - \Gamma^{\mu}_{\kappa \lambda} p^{\kappa} v^{\lambda},  \\
\label{MPD2} \frac{d S^{\mu \nu}}{d \tau} &= 2 p^{[ \mu} v^{\nu ]} - \Gamma^{\mu}_{\kappa \lambda} S^{\kappa \nu} v^{\lambda} - \Gamma^{\nu}_{\kappa \lambda} S^{\mu \kappa} v^{\lambda}, \\
\label{MPD3} v^{\mu} &= \frac{\mathcal{M}}{m^2}\left(p^{\mu} + \frac{2 S^{\mu \nu} R_{\nu \kappa \lambda \rho} p^{\kappa} S^{\lambda \rho}}{4 m^2 + R_{\kappa \lambda \rho \sigma} S^{\kappa \lambda} S^{\rho \sigma}}\right),
\end{align}
where $p^{\mu}$ is the momentum, $S^{\mu \nu}$ is the spin tensor, $m$ is the mass of the particle, which is defined by $m^{2} \equiv -p^{\mu}p_{\mu}$, $v^{\mu}$ is the 4-velocity of the center of mass and is determined by the Tulczyjew-Dixon (TD) spin-supplementary condition (SSC) \cite{Dixon1970},
\be\label{COM1}
p_{\kappa}S^{\kappa \lambda} = 0 
\ee
along with $p^{\mu}v_{\mu} = -\mathcal{M}$. Apart from $S^{\mu \nu}$, another frequently used spin quantity is the spin vector $S^{\mu}$ which can be expressed as 
\be
S^{\mu} = \frac{1}{2 m} \epsilon^{\mu \nu \kappa \lambda} p_{\nu} S_{\kappa \lambda}, 
\ee
where $\epsilon^{\mu \nu \kappa \lambda}$ is the Levi-Civita tensor, and $S^{\mu}$ obviously satisfies
\be\label{COM2}
S^{\kappa} p_{\kappa} = 0.
\ee
Because of  Eq.~(\ref{COM1}) and its anti-symmetry, $S^{\mu \nu}$ only has three linearly independent components and hence can be fully determined by $S^{\mu}$ with
\be
S^{\mu \nu} = \frac{1}{m} \epsilon^{\mu \nu \kappa \lambda} p_{\kappa} S_{\lambda}.
\ee
Meanwhile, we can get a constant of motion if the spacetime admits symmetries described by a Killing vector field $\xi^{\alpha}$ \cite{Rudiger1981},
\be
C = p^{\kappa} \xi_{\kappa} +\frac{1}{2} S^{\kappa \lambda} \nabla_{[\kappa} \xi_{\lambda]}.
\ee
In the case of Kerr spacetime, metric in the usual Boyer-Lindquist coordinate $\left\{t,r,\theta,\phi\right\}$ frame reads
\be
\begin{aligned}
    ds^2 = &-\left(1-\frac{2Mr}{\Sigma}\right)dt^2 - \frac{4Mra\sin^2{\theta}}{\Sigma}dtd\phi + \frac{\Sigma}{\Delta} dr^2 + \Sigma d\theta^2 \\
    &+ \left[\left(r^2+a^2\right)\sin^2{\theta}+\frac{2Mr a^2 \sin^4{\theta}}{\Sigma}\right] d\phi^2 ,
\end{aligned}
\ee
where $M$ is the mass of SMBH, $a$ is its spin parameter and $\Delta \equiv r^2 -2Mr +a^2$, $\Sigma \equiv r^2 + a^2\cos^2{\theta}$.
There exist static and axial symmetry, so we can derive two conservations corresponding to the energy and azimuthal angular momentum of the spinless particle,
\begin{subequations}\label{EandLz}
\begin{align}
E &= -p_{t} + \frac{1}{2} g_{t \kappa , \lambda} S^{\kappa \lambda}, \\
L_{z} &= p_{\phi} - \frac{1}{2} g_{\phi \kappa , \lambda} S^{\kappa \lambda}.
\end{align}
\end{subequations}
And in Kerr spacetime, a quadratic ``conserved'' quantity $Q$, which corresponds to the known Carter constant, can be deduced through linear spin approximation \cite{Rudiger1983,Geoffrey2022}, which takes the form
\be
Q = K - \left(a E - L_z\right)^2, 
\ee
\be
K = Y_{\mu \kappa} Y_{\nu}^{\ \kappa} p^{\mu} p^{\nu} -  m \frac{1}{2}\left(Y_{\mu \nu}^{*}Y^{\mu \nu}\right)_{;\kappa} S^{\kappa} + \frac{2E}{m} Y_{\mu \nu} p^{\mu} S^{\nu},
\ee
where $Y_{\mu \nu}$ is the known Killing-Yano tensor admitted by Kerr spacetime, and $Y_{\mu \nu}^*$ is its dual tensor, which is defined by $Y_{\mu \nu}^* \equiv \frac{1}{2}Y_{\kappa \lambda}\epsilon^{\kappa \lambda}_{\ \ \ \mu \nu}$.

Next, the effect of radiation reactions on orbits should be considered. We discard the terms related to $S^{\mu \nu}$ in calculation as it will involve higher order of mass ratio, so the radiation reaction $F^{\mu}$ for momenta can be simply recovered from the adiabatic radiation fluxes as below,
\begin{subequations}
\begin{align}\label{RR}
&\dot{E} p^{t} = -g_{tt} F^{t} - g_{t\phi} F^{\phi}, \\
&\dot{L_{z}} p^{t} = g_{t\phi} F^{t} + g_{\phi \phi} F^{\phi}, \\
&\dot{Q} p^{t} = 2 g^{2}_{\theta \theta} p^{\theta} F^{\theta} + 2 a^2 E \dot{E} \cos^2\theta + 2 \frac{L_{z} \dot{L_{z}}}{\sin^2\theta} \cos^2\theta, \\
&g_{\mu \nu} u^{\mu} F^{\nu} = 0 ,
\end{align}
\end{subequations}
where the post-Newtonian fluxes $\left\{\dot{E},\dot{L_z},\dot{Q}\right\}$ in terms of the orbit elements $\left\{p,e,\iota\right\}$ we adopted are those in Ref.~\cite{Gair2006}. During inspirals, semi-latus rectum $p$, eccentricity $e$ and ``inclination angle'' $\iota$ are calculated through
\begin{subequations}
\begin{align}
&p = \frac{r_\text{max} \times r_\text{min}}{r_\text{max} + r_\text{min}}, \\
&e = \frac{r_\text{max} - r_\text{min}}{r_\text{max} + r_\text{min}}, \\
&\tan^2 \iota = \frac{Q}{L^{2}_{z}}.
\end{align}
\end{subequations}
The fact that we ignore the explicit parts about $S^{\mu \nu}$ in radiation reaction formulas doesn't mean all radiation effects of $S^{\mu \nu}$ are discarded, as the appearance of spin will perturb the orbit elements with same initial conditions. That is to say, we mainly focus on the impact of the secondary spin on the orbital motion instead of the GW emission, because measurement precision of the spin is the main concern in this paper and it should be naturally improved when more effects are included. With those in hand, the approximate inspirals of the sBH are calculated from Eq.~(\ref{MPD2}) and Eq.~(\ref{MPD3}) with
\be\label{MPD4}
\frac{d p^{\mu}}{d \tau} = -\frac{1}{2} S^{\kappa \lambda} v^{\rho} R^{\mu}_{\ \rho \kappa \lambda} - \Gamma^{\mu}_{\kappa \lambda} p^{\kappa} v^{\lambda} + F^{\mu},
\ee
as we will integrate these equations numerically by the Runge-Kutta-Fehlberg method to obtain the trajectory of the sBH.

For evolving the equations of motion above [Eqs.~(\ref{MPD2},~\ref{MPD3},\ref{MPD4})], we need to specify the initial values 
of dynamical variables $\left\{S^{\mu}, p^{\mu}, x^{\mu} \right\}$. In 
the case of generic misaligned and eccentric orbits, we choose
\be\label{param_in}
\lambda^{i}_{\rm in} = \left\{m, M, a, s, S^{r}_0, S^{\phi}_0, p_0, e_0, \iota_0, r_0, \theta_0\right\}
\ee
as the initial intrinsic parameters, while initial time $t_0$ and initial phase in the azimuthal direction $\phi_0$ are set to 0, and $s$ is the dimensionless spin parameter of the sBH which is defined as $s \equiv \frac{1}{m}\sqrt{S^{\mu}S_{\mu}}$. Then $\left\{ p^{\mu}_0 \right\}$ are resolved by the relationship between orbit elements and momenta for geodesic orbits \cite{Schmidt2002}, and $\left\{ S^{t}_0, S^{\theta}_0 \right\}$ are resolved by Eq.~(\ref{COM2}). Apart from $s$, frequently mentioned spin parameters are its projection onto the orbital angular momentum $s_{||}$ and another component $s_{\bot}$, which satisfy $s^2 = s^2_{||} + s^2_{\bot}$. For resolving the EMRI formation, which will be discussed in the next section, the sBH should have no preference for the direction of the spin, so we focus on the total magnitude $s$ instead of its projections.
%In order to connect with other spinless EMRI models, we choose the spinless particle's relationship between $\left\{p, e, \iota\right\}$ and $p^{\mu}$, which has an analytic form \cite{Schmidt2002}, to determine initial momenta  So our initial orbit elements are not true ones, but rather those in the spinless case. $S^t_0$ and $S^{\theta}_0$ are recovered by Eq.~(\ref{COM2}) combined with $S^r_0$, $S^{\phi}_0$, $s$ and the initial momenta.

After finishing integration of the sBH's inspiral orbit, we can calculate the transverse-traceless (TT) gravitational waveform through quadrupole approximation by
\begin{align}
\overline{h}^{TT}_{ij}(t) &= \frac{2}{D} \Lambda_{ij,kl}\left[\Ddot{I}^{kl}(t^{'})\right]_{t^{'}=t-R}, \\
\Lambda_{ij,kl} &= P_{ik} P_{jl} - \frac{1}{2} P_{ij} P_{kl}, \\
I^{kl} &= m x^{k} x^{l},
\end{align}
where $D$ is the distance to the EMRI source, $P_{ij}$ is the projection operator which is defined as
\be
P_{ij} \equiv \delta_{ij} - n_i n_j,
\ee
and vector $\hat{n}$ is the direction of GW propagation.

It is necessary to transform waveforms from the source frame to the solar-system barycenter (SSB) frame. Let $\hat{A}$ be the direction of the SMBH spin and the $z$ axis of the source frame, $\hat{N}$ be the direction pointing to the source system . Then let $\pm \hat{N} \times \hat{A}$ be the principal `$+$' direction and an axis, which is rotated around $-\hat{N}$ counterclockwise from it by $45^\circ$, be the principal `$\times$' direction. The polarization of GW is then expressed by
\be
\begin{aligned}
h_+ = &\left(\cos^2{\Phi} - \sin^2{\Phi}\cos^2{\Theta} \right) \frac{\Ddot{I}^{xx}}{D} +  \\
&\left(\sin^2{\Phi} - \cos^2{\Phi}\cos^2{\Theta} \right) \frac{\Ddot{I}^{yy}}{D} -  \\
&\sin^2{\Theta} \frac{\Ddot{I}^{zz}}{D} - \sin{\Phi} \cos{\Phi} \left(1 + \cos^2{\Theta}\right) \frac{2\Ddot{I}^{xy}}{D} + \\
&\sin{\Phi}\sin{\Theta}\cos{\Theta}\frac{2\Ddot{I}^{xz}}{D} + \cos{\Phi}\sin{\Theta}\cos{\Theta}\frac{2\Ddot{I}^{yz}}{D},
\end{aligned}
\ee
\be
\begin{aligned}
h_{\times} =& - \sin{2\Phi}\cos{\Theta}\frac{\Ddot{I}^{xx}-\Ddot{I}^{yy}}{D} - \cos{\Theta}\cos{2\Phi}\frac{2\Ddot{I}^{xy}}{D} \\
&+\sin{\Theta}\cos{\Phi}\frac{2\Ddot{I}^{xz}}{D} - \sin{\Theta}\sin{\Phi}\frac{2\Ddot{I}^{yz}}{D},
\end{aligned}
\ee
where $\Theta$ is the angle between $\hat{N}$ and $\hat{A}$, $\Phi$ is the initial azimuthal angle of sBH with respect to $\hat{N} \times \hat{A}$. These source-frame waveforms are then further transformed to SSB frame by
\be
h_{+,\text{SSB}} =  h_{+}\cos{2\psi} - h_{\times}\sin{2\psi}, 
\ee
\be
h_{\times,\text{SSB}} = h_{+}\sin{2\psi} + h_{\times}\cos{2\psi},
\ee
where the polarization angle $\psi$ is the angle between `+' direction and $\hat{z}$ axis of the SSB frame and can be calculated by
\be
\psi = \arctan\left(\frac{\hat{A}\cdot\hat{z} - \left(\hat{A}\cdot\hat{N}\right)\left(\hat{z}\cdot\hat{N}\right)}{\hat{N}\cdot\left(\hat{A}\times\hat{z}\right)}\right).
\ee
So, we set extrinsic system parameters as
\be
\lambda^i_{ex} = \left\{T,\theta_{A},\phi_{A},\theta_{N},\phi_{N},\Phi,D\right\}
\ee
and
\begin{equation}
\boldsymbol{\lambda} \equiv \left(\lambda^1,\dots,\lambda^{18}\right) 
= \left(\lambda^1_{\rm in},\dots,\lambda^{11}_{\rm in},\lambda^1_{\rm ex},\dots,\lambda^7_{\rm ex}\right),
\end{equation}
where $T$ is the duration of gravitational waveform, $\theta_S$ and $\phi_S$ represent the direction of $\hat{A}$ in the SSB frame, $\theta_N$ and $\phi_N$ represent $\hat{N}$.

LISA and Taji will detect the GW signal with an interferometric measurement of differential optical pathlength modulation along the arms between three free-falling test masses inside spacecraft, i.e. the relative frequency shift $\left\{y_{12}(t),y_{21}(t),y_{13}(t),y_{31}(t),y_{23}(t),y_{32}(t)\right\}$ where the lower indices mark the spacecraft. A change of light-travel distance caused by the GW should be computed by the projection of the above SSB-frame waveforms onto the detector arms, and the response function depends on not only the frequency of signals but also the time due to the typically long duration of EMRI waveforms. Moreover, the laser noise will drown out the GW signal with realistically unequal arms, thus time-delay interferometry (TDI) technology was proposed by Ref.~\cite{Tinto1999,Tinto2002,Tinto2004,Tinto2014}. In short, the key point of TDI is the linear combination of time-delay GW projections in order to construct equivalent equal-length arms and suppress the laser noise. For the second-generation TDI designed for evolving length of arms, the Michelson combination, $X$, is given by \cite{Tinto2004}
\be
\begin{aligned}
X = &y_{13} + \mathcal{D}_{13} y_{31} + \mathcal{D}_{131} y_{12} + \mathcal{D}_{1312} y_{21} \\
& - \left(y_{12} + \mathcal{D}_{12} y_{21} + \mathcal{D}_{121} y_{13} + \mathcal{D}_{1213} y_{31}\right) \\
& + \mathcal{D}_{13121} y_{12} + \mathcal{D}_{131212} y_{21} + \mathcal{D}_{1312121} y_{13} + \mathcal{D}_{13121213} y_{31} \\
& - \left( \mathcal{D}_{12131} y_{13} + \mathcal{D}_{121313} y_{31} + \mathcal{D}_{1213131} y_{12} + \mathcal{D}_{12131312} y_{21} \right),
\end{aligned}
\ee
where the delay operator is defined by
\be
\mathcal{D}_{i_1 i_2 ... i_n} x(t) \equiv x\left(t-\sum_{k=1}^{n-1} L_{i_k i_{k+1}}(t)\right),
\ee
and $L_{ij}(t)$ is the propagation time at reception time $t$. Applying cyclic permutation of the lower index, we can obtain $Y$ and $Z$. These three combinations still have correlated noise properties, and an uncorrelated set of observables $\left\{A,E,T\right\}$ can be obtained by
\begin{subequations}
\begin{align}
A& = \frac{1}{\sqrt{2}} \left(Z - X\right), \\
E& = \frac{1}{\sqrt{6}} \left(X - 2Y + Z\right), \\
T& = \frac{1}{\sqrt{3}} \left(X +Y + Z\right).
\end{align}
\end{subequations}

In this paper, we plug the above SSB-frame waveforms into the \href{https://github.com/mikekatz04/lisa-on-gpu}{fastlisaresponse} framework \cite{Katz2022} along with the orbit information of Taiji, to compute the second-generation TDI response $\left\{h_{A},h_{E},h_{T}\right\}$. We label corresponding power spectral density (PSD) of the noise for each TDI channel as $\left\{S_{n,A},S_{n,E},S_{n,T}\right\}$, and we will apply the PSD of Taiji detector to the calculation in following sections.

\begin{figure}
    \centering
    \includegraphics[width=1.\linewidth]{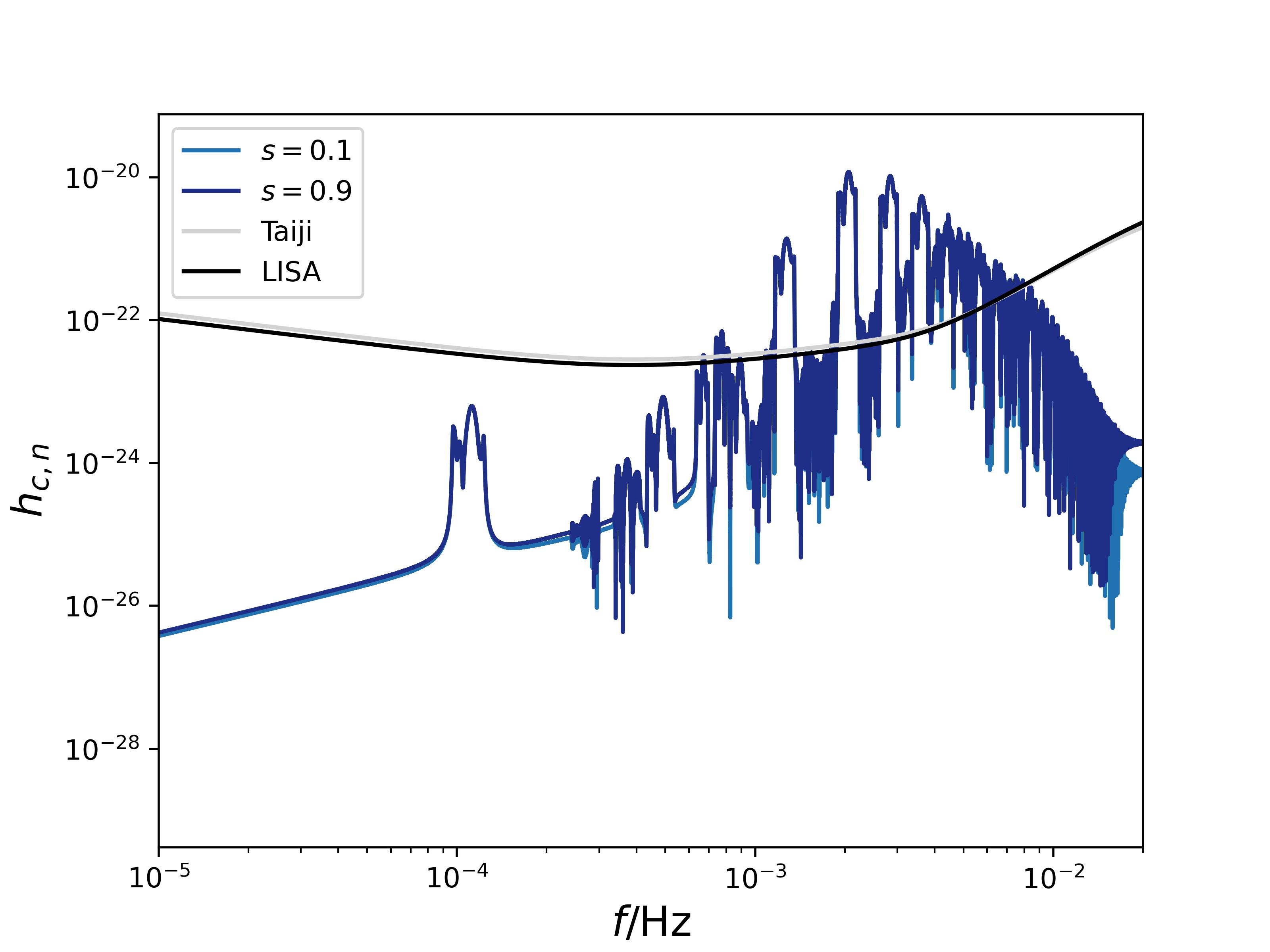}
    \caption{The characteristic strain $h_c$ for $s=0.1$ (blue curve) and $s=0.9$ (purple curve), and the noise amplitude $h_n$ of Taiji (light gray curve) and LISA (black curve).}
    \label{fig:wave}
\end{figure}

As an example, we plot frequency-domain waveforms for $s=0.1$ and $s=0.9$ in Fig.~\ref{fig:wave} through the A-channel characteristic strain $h_c$ which is defined as \cite{Moore2015}
\be
h_c(f) = 2 f \lvert \Tilde{h}_A(f) \lvert
\ee
where hat ``$\sim$'' labels the Fourier transformation of signal, and along with the noise amplitude $h_n$ which is 
\be
h_n(f) = \sqrt{fS_{n,A}(f)}.
\ee
Other intrinsic parameters are set as $m = 10 M_\odot$, $M = 1 \times 10^6 M_\odot$, $a = 0.99M$, $S^r_0=S^\phi_0=0$, $p_0=10M$, $e_0=0.2$, $\iota_0=0.5$, $r_0=p_0$, $\theta_0=\frac{\pi}{2}$ (where the subscript ``0" means the initial time), and extrinsic parameters are set as $T=1 \text{yr}$, $\theta_A=\frac{\pi}{2}$, $\phi_A=\frac{\pi}{5}$, $\theta_N=\frac{\pi}{3}$, $\phi_N=\frac{\pi}{4}$, $\Phi=\frac{\pi}{2}$, $D=1$ Gpc. Note that, although we take initial radial and azimuthal components of $S^{\mu}$ to be 0, they will become nonzero as a result of precession for generic misaligned orbits.
%so this condition can still be viewed as no preference of orbit for the secondary spin. 

\subsection{Fisher forecast}\label{subsection:precision}
Assuming the detector noise is stationary and Gaussian, the likelihood of detecting data $\mathbf{d}$ 
with a signal $\mathbf{h}(\boldsymbol{\lambda_0})$
is 
\be
\mathcal{L}\left(\mathbf{d} \mid \boldsymbol{\lambda_0}\right) \propto e^{-\left(\mathbf{d} - \mathbf{h}(\boldsymbol{\lambda_0}) \mid \mathbf{d} - \mathbf{h}(\boldsymbol{\lambda_0})\right)/2}\ ,
\ee
where the inner product of two signals, $\mathbf{a}(t)$ and $\mathbf{b}(t)$, is defined as \cite{Cutler1994}
\be
\left(\mathbf{a} \mid \mathbf{b}\right) = \sum_{i = A,E,T} 2\int^{\infty}_{0}\frac{\Tilde{a}^{*}_{i}(f) \Tilde{b}_{i}(f) + \Tilde{a}_{i}(f) \Tilde{b}^{*}_{i}(f)}{S_{n,i}(f)} df,
\ee
and the upper index ``$*$'' labels the complex conjugate.
And the signal-to-noise ratio (SNR) $\rho$ for a true waveform $\mathbf{h}$ is defined as
\be
\rho \equiv \sqrt{\left(\mathbf{h} \mid \mathbf{h}\right)}.
\ee
\textcolor{red}{
To estimate the system parameter, one should fully explore the posterior distribution by Bayesian analysis (e.g., Markov-chain Monte Carlo). But for EMRI waveforms, to search across the large parameter space requires template waveforms that can be generated in less than a second, which remains unachieved for the generic orbit and lies beyond the study's primary objectives.
So, in this paper, we use the Fisher Information Matrix (FIM) approximation as a practical method to evaluate the measurement precision, which is widely adopted in GW research, although with known inadequacies \cite{Rodriguez2013,Vallisneri2008}. 
}
For large SNR,  the likelihood $\mathcal{L}$ is approximately a Gaussian distribution centered on the true physical values \cite{Vallisneri2008},
\be
\mathcal{L}\left(\boldsymbol{\lambda} \mid \mathbf{d}\right) \sim e^{-\Gamma_{ij}\left(\lambda^i-\lambda^i_0\right)\left(\lambda^j-\lambda^j_0\right)/2},
\ee
where 
\be\label{Eq:fisher}
\Gamma_{ij} = \left(\frac{\partial \mathbf{h}}{\partial \lambda^i} \mid \frac{\partial \mathbf{h}}{\partial \lambda^j}\right)\ ,
\ee
is the FIM.
So, we define the measurement precision through the covariance matrix $\Gamma^{-1}$ by
\be
\Delta \lambda^i \equiv \sqrt{\text{Cov}\left(\lambda^i, \lambda^i\right)} = \sqrt{\left(\Gamma^{-1}\right)_{ii}}.
\ee

\begin{figure}
    \centering
    \includegraphics[width=1.1\linewidth]{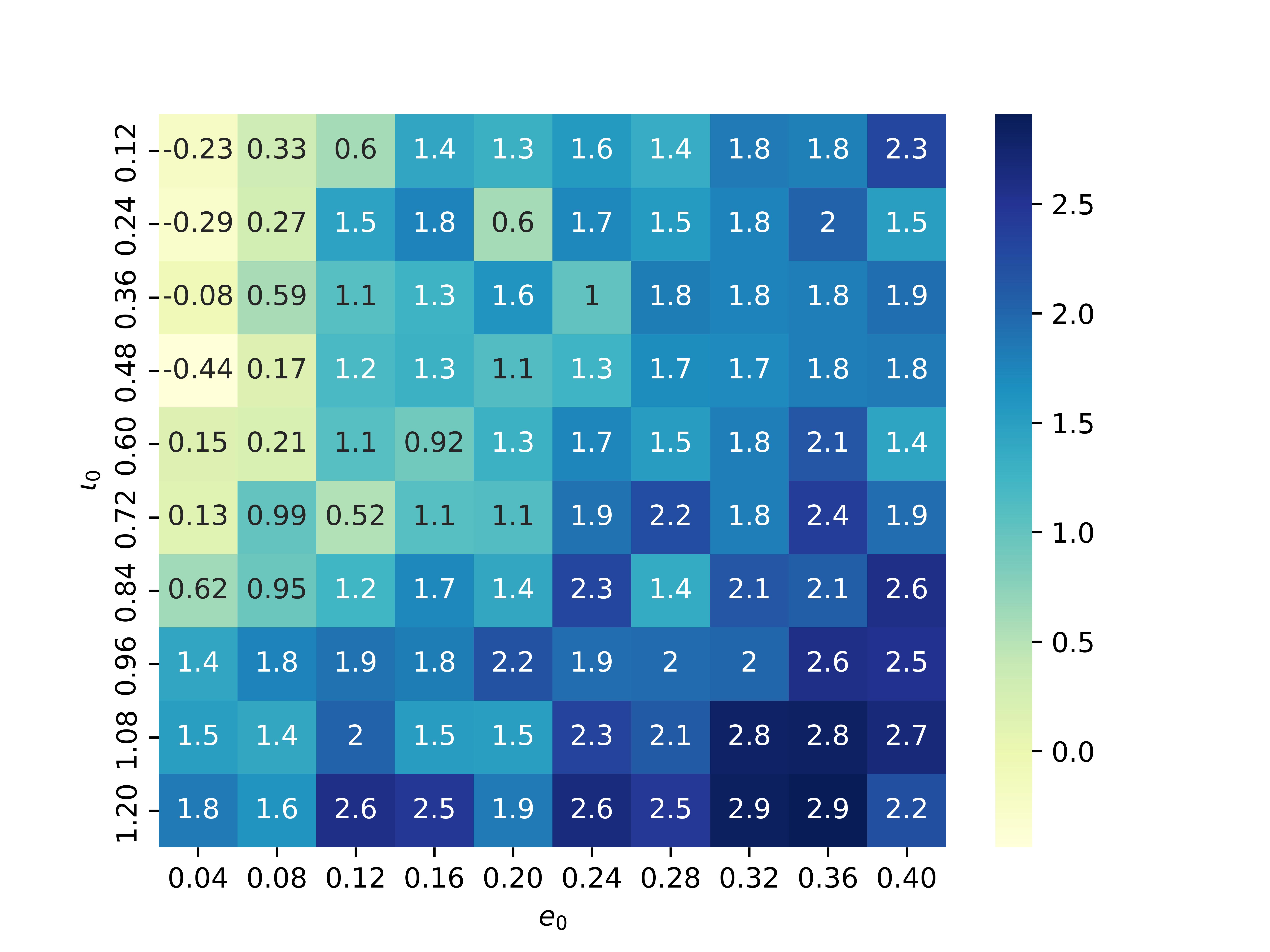}
    \caption{The dependence of the secondary spin measurement precision $-\log_{10}(\Delta s/s)$ on initial values of inclination angle $\iota_0$ and eccentricity $e_0$.  }
    \label{fig:heatmap}
\end{figure}

Fig.~\ref{fig:heatmap} shows how the relative measurement precision of the secondary spin, $\Delta s/s$, varies with the orbital eccentricity $e_0$ and inclination angle $\iota_0$, and other parameters are set as the same as in Fig.~\ref{fig:wave} except $s=0.5$ and the luminosity distance $D$ is adjusted so that SNR $\rho=20$. There is a clear dependence of the measurement precision on $e_0$ as well as $\iota_0$. For a near-spherical initial orbit where $e_0=0.04$, if $\iota_0$ is lower than $\frac{\pi}{4}$, $\Delta s$ is higher than 0.35 and relative error $\Delta s/s>70\%$. If the orbit is further near-equatorial where $\iota_0=0.12$, $\Delta s$ becomes larger than 1, so we will get little information about $s$ in this situation. 
The measurement precision improves significantly in the case of large $e_0$ and/or $\iota_0$. 

To figure out the reason for this tendency, we concentrate on the linear term of the secondary spin and neglect the higher order terms, then the covariant MPD equation can be approximated by
\be\label{app_MPD1}
\frac{D p^{\mu}}{d\tau} = - \frac{1}{2m} R^{\mu}_{\ \rho \kappa \lambda} p^{\rho} S^{\kappa \lambda},
\ee
\be
\frac{D S^{\mu \nu}}{d\tau} = 0,
\ee
\be
m v^{\mu} = p^{\mu}.
\ee
The effects of the secondary spin enter the dynamics via its coupling with momentum  $p^{\mu}$ as formulated in Eq.~(\ref{app_MPD1}). If the orbit is close to equatorial, $p^{\theta}$ is bound around 0. As a result, the terms related to $p^{\theta}$ in the extra force $R^{\mu}_{\ \rho \kappa \lambda} p^{\rho} S^{\kappa \lambda}$ arising from the secondary spin will be negligible too. 
%With fewer non-zero terms of $s$ included when integrating, $\Gamma_{ss}$ should certainly decrease. 
The same happens with near-spherical orbits and $p^r$. More specifically, we choose equatorial-aligned orbit as an example, in which $Dp^{\theta}/d\tau=0$ requires $S^{\mu} = (0,0,S^{\theta},0)$, and combine with $p^{\mu} = (p^t,p^r,0,p^{\phi})$, then
\be\label{app_MPD2}
\begin{aligned}
\frac{D p^{\phi}}{d\tau} &= -\frac{1}{2m} R^{\phi}_{\ \rho \kappa \lambda} \epsilon^{\kappa \lambda}_{\ \ \ \nu \theta} p^{\rho} p^{\nu} S^{\theta} \\
&= -\frac{1}{m} R^{\phi}_{* \rho \nu \theta} p^{\rho} p^{\nu} S^{\theta} \\
&= -\frac{1}{m} \left(R^{\phi}_{* t r \theta} p^t + R^{\phi}_{* r t \theta} p^t 
+ R^{\phi}_{* r \phi \theta} p^{\phi}+ R^{\phi}_{* \phi r \theta} p^{\phi}\right) S^{\theta} p^r,
\end{aligned}
\ee
where dual Riemann tensor $R^{\alpha}_{* \beta \gamma \delta} \equiv \frac{1}{2} R^{\alpha}_{\ \beta \mu \nu} \epsilon^{\mu \nu}_{\ \ \ \gamma \delta}$ and we have included the non-zero components only. When $p^{r} = 0$, which corresponds to circular orbits, $D p^{\phi}/d\tau$ should be 0 as a natural result of its constant azimuthal frequency \cite{Han2010}. 
%{\bf That is to say, for circular orbits, the secondary spin $s$ only perturbs the initial momenta with the same energy and angular momentum as it appears in Eq.~(\ref{EandLz}),} 
That is to say, for circular orbits, the momentum evolution and the secondary spin evolution are decoupled,
so the waveform should contain little information about $s$ in this situation. This conclusion is consistent with the previous detailed studies \cite{Gair2011,Piovano2021}. However, in the case of an eccentric orbit, $p^r$ varies and deviates from zero, then the coupling terms appearing in Eq.~(\ref{app_MPD2}) become nonzero. The effect of the secondary spin is larger for more eccentric orbits as the coupling is proportional to $p^r$. Furthermore, for generic orbits, more coupling terms will be released, as $p^{\theta}$ is non-zero and $S^{\mu}$ has more non-zero components, so it is possible to measure the secondary spin with a reasonable accuracy.

% This conclusion can be confirmed by the overlap result in Fig.~\ref{fig:overlap} where $e_0 = \iota_0 = 0.01$ for the near-circular orbit and $e_0=0.2, \iota_0=0.5$ for the generic orbit, other parameters except the varied secondary spins keep the same with Fig.~\ref{fig:wave}. The overlap, which is defined as
% \be
% \text{overlap} \equiv \frac{\left(\mathbf{a} \mid \mathbf{b}\right)}{\sqrt{\left(\mathbf{a} \mid \mathbf{a}\right)\left(\mathbf{b} \mid \mathbf{b}\right)}},
% \ee
% represents the ``goodness-of-fit'', and we label...\textbf{maybe this overlap figure is unnecessary}.
% \begin{figure}
%     \centering
%     \includegraphics[width=1.\linewidth]{pic/overlap.jpg}
%     \caption{Overlap between the spinless case and varied secondary spins ones for generic and near-circular orbits.}
%     \label{fig:overlap}
% \end{figure}

\begin{figure}
    \centering
    \includegraphics[width=1.\linewidth]{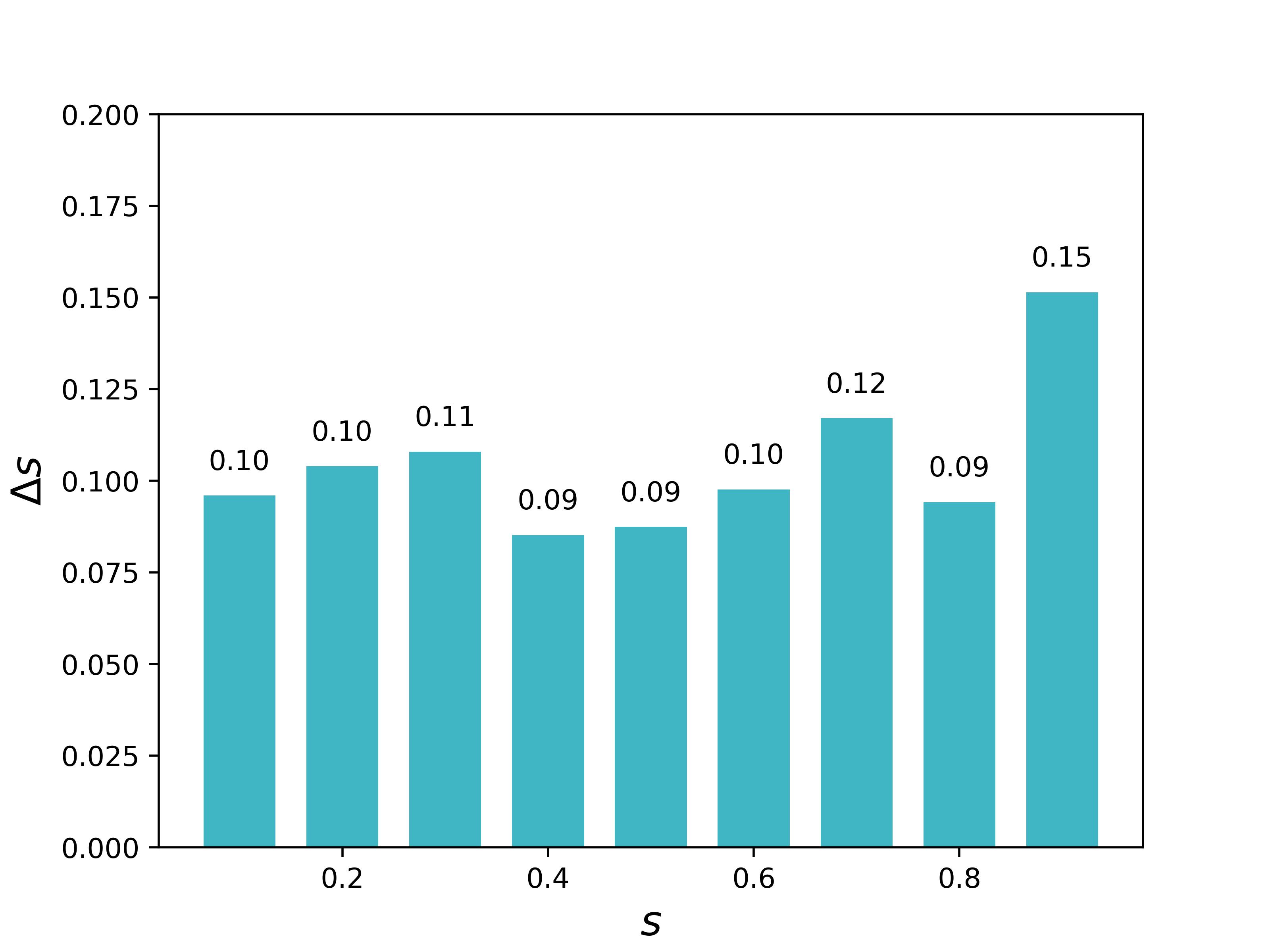}
    \caption{The expected measurement uncertainties $\Delta s$ of secondary spins  for different $s$.} %\todo{which is the orbital parameters cui: already mentioned in the main body of paper}
    \label{fig:bar}
\end{figure}

%Also, we present the measurement precision 
In Fig.~\ref{fig:bar}, we show the measurement precision for a range of secondary spin $s$ (other parameters are the same to in Fig.~\ref{fig:wave} and $D$ is also adjusted so that $\rho=20$). The uncertainty $\Delta s$ roughly stays the same ($\sim 0.1$) for different
secondary spins $s$. This is because the effect of the secondary spin in the EMRI waveform is dominated by its linear order, i.e., $\partial\mathbf{h}/\partial s$
is approximately independent of the spin magnitude $s$.
%which can be understood from the fact that $s$ in the leading linear-spin terms should disappear when calculating the deviation in Eq.~(\ref{Eq:fisher}). %But note that, with unchanged $\Delta s$, the relative precision would decrease as the spin magnitude drops. 
%The relative precision roughly decreases with the secondary spin, but the main reason for this tendency is the decrease of $s$ itself, while $\Delta s$ doesn't change greatly. It can be understood from the fact that $s$ in the leading linear-spin terms should disappear when computing the deviation in Eq.~\ref{Eq:fisher}, thus the Fisher information matrix term $\Gamma_{ss}$ would stay the same for the linear approximation with different secondary spins. 
%It can be seen clearly that when $p^{r} = 0$, which corresponds to circular orbit, the linear effects disappear and only higher order terms of 2nd spin work, so waveforms contain very little information about $s$ in this situation. And with eccentricity increasing, $p^r$ starts to evolve then $\Gamma_{ss}$ gets improved.
%In a word, for circular orbits, the spin only give a constant change on the orbital frequency but without the coupling between the secondary spin and the momentum, so it is hard to measure the secondary spin in this situation from GWs, this coincides with the previous researches (\cite{gair}). However, for the planar and eccentric case, the secondary spin will couple with the momentum.  Moreover, in the generic orbits, not the magnitude of spin coupling with momentum but also the direction (i.e., the precession) . Therefore for eccentric and especially the generic orbits, it is possible to measure the secondary spin with a reasonable accuracy (see Fig. ...) . 

\section{Secondary spins and EMRI formation channels}

\subsection{Modeling secondary spins of dry and Hills EMRIs}\label{subsec:spin model}

As briefly summarized in the Introduction, three main EMRI formation channels have been well established in the literature: dry channel \cite{Hopman2005,Preto2010,Bar-Or2016,Babak2017,Amaro2018,Broggi2022},  Hills channel \cite{Miller2005,Raveh2021} and  wet  channel \cite{Sigl2007,Levin2007,Pan2021prd,Pan2021b,Pan2021,Pan2022,Derdzinski2023}.
A unique signature of wet EMRIs is the low orbital eccentricity in the mHz band.
Dry and Hills EMRIs are quite similar in the orbital parameters, therefore how to distinguish them is a remain-to-be-answered question.
In the remainder of this section, we will first parametrize the distribution of secondary 
spins of dry EMRIs and Hills EMRIs, then test whether we can measure the contribution to future EMRI detections of each channel via an injection and inference experiment.

Dry EMRIs keep the birth spins of sBHs formed from individual massive stars, 
which are of low values as inferred from LVK binary BH merger events \cite{LIGOScientific:2018jsj,Roulet:2018jbe,Fuller2019,KAGRA:2021duu} and we parameterize the secondary spin distribution in a (semi-)Gaussian form
\be 
P_{1}(s) = N(\sigma_1) e^{-\frac{s^2}{2\sigma_1^2}}\ ,
\ee 
where $N(\sigma_1)$ is a normalization factor.

For comparison, Hills EMRIs keep the birth spins of sBHs formed from binary massive stars.
The spins of the first-born sBHs approximately follow the same distribution $P_1(s)$ of isolated sBHs, 
while the spins of the second-born sBHs are largely determined by the tidal spin up during the pre-collapse Helium core phase, thereby follow a different distribution $P_2(s)$, which is commonly believed to be a roughly bimodal distribution \cite{Zaldarriaga:2017qkw,Bavera:2020inc,Mandel:2020lhv,Adamcewicz:2023szp}. 
In the evolution track of a massive star binary, the more massive star evolves faster and collapses into a BH first, 
and the lighter one later expands largely during its red giant phase, forming an common envelope. 
Within a dynamical timescale, the binary separation shrinks by orders of magnitude and the common envelope is ejected \cite[see e.g., Ref.~][for explicit binary evolution examples]{Olejak2020}.
In this naked Helium core + first-born BH stage, the He-core can be tidally spun up by the companion BH.
In the case of wide binaries, the tidal effect is weak, and the lifetime time of the He-core  $T_{\rm He}$ is shorter than the tidal synchronization timescale $T_{\rm sync}$.
As a result, the He-core has not been tidally locked at collapse and the spin of the second-born BH scales as 
\cite{Zahn1977}
\be 
s_2\propto \omega_{\rm He}\approx \frac{T_{\rm He}}{T_{\rm sync}}\omega_{\rm b} \propto A_{\rm b}^{-15/2}\quad ({\rm for}\ T_{\rm He} < T_{\rm sync}),
\ee 
where $A_{\rm b}$ is the He-core+BH binary separation, $\omega_{\rm b}$ is the binary orbital angular frequency,
$\omega_{\rm He}$ is the angular frequency of the He-core.
Assuming a logUniform distribution of the binary separation $A_{\rm b}$, with a lower limit $A_{\rm b, min}$ set by the binary contact limit and 
an upper limit set by the binary orbital evolution during the common envelope phase $A_{\rm b, max}$, 
we expect a distribution of second-born BH spins in the form of 
\be 
P_{\rm non-sync}(s_2)={\rm logUniform}[s_{2,\rm min}, 1]\ ,
\ee 
where $s_{2, \rm min}$ is the second-born BH spin in the widest binary ($A_{\rm b, max}$) after the common envelope phase.

The He-core can be spun up until being tidally locked in the case of a tight binary with $A_{\rm b, min}< A_{\rm b} < A_{\rm b, sync}$, 
where $A_{\rm b, sync}$ is the critical binary separation for the He-core getting synchronized before collapse.
In this case, the spin of the second-born BH from the He-core collapse $s_2$ scales as \cite{Bavera:2020inc}
\be
s_2 \propto \omega_{\rm He} \approx  \omega_{\rm b}\propto  A_{\rm b}^{-3/2}\quad ({\rm for}\ T_{\rm He} > T_{\rm sync}) ,
\ee 
and is capped by the maximum spin $1$. As we will see that a large fraction of tight binaries produce fast spinning second-born BHs,
and the distribution is approximated by a Gaussian distribution peaked at $1$,
%Assuming a typical lifetime $T_{\rm He}=0.25$ Gyr for  He-cores of $\approx 10 M_\odot$, 
%we find the critical binary separation at which $T_{\rm He} = T_{\rm sync}$ as 
%\be 
%A_{\rm b, crit}  \approx 10^{6.6} {\rm km} \left( \frac{m_1}{10 M_\odot}\right)^{1/4} \left( \frac{m_2}{10 M_\odot}\right)^{1/4} \left( \frac{m_1+m_2}{20 M_\odot}\right)^{1/4}\ ,
%\ee 
\be 
P_{\rm sync}(s_2)=N(\sigma_2) e^{-\frac{(s_2-1)^2}{2\sigma_2^2}}\ .
\ee 
Therefore, we expect a bimodal distribution of spins of second-born BHs in the form of 
\be 
P_2(s)=\eta P_{\rm non-sync}(s) + (1-\eta) P_{\rm sync}(s)\ ,
\ee  
where $\eta$ quantifies the weight of the low-spin component, or equivalently the fraction of wide binaries. 

Taking the binary evolution simulation results in Ref.~\cite{Bavera:2020inc} as a reference, we have $\log_{10}(s_{2, \rm min})\approx -4$, 
typical lifetime of $\sim 10 M_\odot$ He-core as $T_{\rm He}\approx 0.25$ Gyr 
and the merger timescale of the widest binary $T_{\rm merger}(A_{\rm b, max})\approx 15$ Gyr ,
which yields
\be 
\begin{aligned}
    A_{\rm b, max} & \approx 10^{7.1}\ {\rm km} \left(\frac{m_1}{10 M_\odot}\right)^{1/4} \left(\frac{m_2}{10 M_\odot}\right)^{1/4} \left(\frac{m_1+m_2}{20 M_\odot}\right)^{1/4}\ , \\
    A_{\rm b, sync} &\approx 10^{6.6}\ {\rm km} \left(\frac{m_1}{10 M_\odot}\right)^{1/4} \left(\frac{m_2}{10 M_\odot}\right)^{1/4} \left(\frac{m_1+m_2}{20 M_\odot}\right)^{1/4}\ ,
\end{aligned}
\ee 
where $m_1$ and $m_2$ are the masses of the first born and second born BHs, respectively.
In combination with the minimal binary separation, the He-core radius \cite{Kippenhan1990book}
\be 
A_{\rm b, min} = R_{\rm He} \approx 10^{5.8} \ {\rm km} \left(\frac{m_{\rm He}}{10 M_\odot}\right)^{0.7} ,
\ee 
we obtain the fiducial value  $\eta\approx0.4$, where we have used the assumption of 
logUniform distribution of the binary separation, $\eta = \log(A_{\rm b, max}/A_{\rm b, sync})/\log(A_{\rm b, max}/A_{\rm b, min})$.

To summarize, we expect low secondary spins of dry EMRIs with probability distribution 
\be\label{dry_spin} 
P_{\rm dry}(s) = P_1(s)\ ,
\ee 
and a mixture of low and high secondary spins of Hills EMRIs with probability distribution 
\be \label{Hills_spin}
P_{\rm Hills}(s) = \frac{P_1(s)+P_2(s)}{2}  \ ,
\ee 
where we have assumed an equal probability of first and second born BHs captured by the SMBH after binary disruptions.
In general, each channel contributes a fraction of EMRIs and the total distribution is formulated as 
\be \label{total_spin}
P(s) = f_{\rm dry}P_{\rm dry}(s)+ (1-f_{\rm dry})P_{\rm Hills}(s) \ ,
\ee 
where the fraction  of dry EMRIs $f_{\rm dry}$ is highly uncertain to predict from first principles.
In this model, high-spin secondary BHs are a signature of Hills EMRIs.
In the remainder of this section, we will quantify how accurately we can measure the branch ratios 
of the two channels from a number of EMRI detections.

\subsection{Population inference basics}\label{subsection:infer basic}

The hierarchical Bayesian method has been widely used in population inference of stellar mass binary mergers 
detected by ground-based detectors LIGO/Virgo/KAGRA. We will use the same method to constrain the EMRI population properties in this subsection.
From loud events that can be individually detected $\{ \mathbf{d}_i\} \ (i=1,...,N_{\rm det})$, one can infer the total number of all events $N_{\rm tot}$ and the population parameters $\mathbf{\Lambda}$, with the population likelihood \cite{LVCO1O2,LVKCO3}
\begin{equation}
    \label{eq:PopLikelihood}
    \mathcal{L}(\{ \mathbf{d}_i\} | \mathbf{\Lambda}, N_{\rm tot}) \propto N_{\rm tot}^{N_{\rm det}} e^{-N_{\rm tot} \xi(\mathbf{\Lambda})} \prod_{i=1}^{N_{\rm det}} \int \mathcal{L}\left( \mathbf{d}_i | \boldsymbol{\lambda}\right) p_{\rm pop}(\boldsymbol{\lambda} | \mathbf{\Lambda}) \dd \boldsymbol{\lambda} \ .
\end{equation}
The $\xi(\mathbf{\Lambda})$ term represents the fraction of detectable EMRIs in the population $p_{\rm pop}(\boldsymbol{\lambda} | \mathbf{\Lambda})$, where $\mathbf{\Lambda}$ is the EMRI population parameters, $\boldsymbol{\lambda}$ is the EMRI waveform parameters,
and 
$\mathcal{L}\left( \mathbf{d}_i | \boldsymbol{\lambda}\right)$ is the likelihood of seeing data $\mathbf{d}_i$ in GW detectors
from an EMRI with parameters $\boldsymbol{\lambda}$.

In our case, we have no plan to explore the population properties of all the EMRI waveform parameters $\boldsymbol{\lambda}$, most of which are not
relevant in distinguishing dry and Hills EMRIs. For this purpose,
only the secondary spin $s$ matters, which is not expected to largely affect the EMRI signal strength or the selection function $\xi(\mathbf{\Lambda})$, therefore 
the population likelihood simplifies as
\begin{equation}
    \label{eq:PopLikelihood}
    \mathcal{L}(\{ \mathbf{d}_i\} | \mathbf{\Lambda}) \propto \prod_{i=1}^{N_{\rm det}} \int \mathcal{L}\left( \mathbf{d}_i | s\right) P(s | \mathbf{\Lambda}) \dd s \ ,
\end{equation}
where $\mathbf{\Lambda}=\{\sigma_1,\log_{10} s_{2,\text{min}},\eta,\sigma_2,f_\text{dry}\}$ are the population model parameters of the secondary spins [Eq.~(\ref{total_spin})],
$\mathcal{L}\left( \mathbf{d}_i | s\right)$ is the likelihood marginalized over all other waveform model parameters and is approximated as
\be\label{eq:detectlikelihood}
\mathcal{L}\left( \mathbf{d} | s\right) \propto \exp\left\{-\frac{(s-s^{\rm inj})^2}{2(\Delta s)^2}\right\}\ ,  
\ee 
with the injection secondary spin $s^{\rm inj}$ and the expected measurement uncertainty $\Delta s$ as inferred from Fisher forecasts in the previous section.

\subsection{Secondary spins of EMRIs: injection and inference}\label{subsection:inference result}

As a fiducial EMRI population,  we consider a redshift-independent massive black hole (MBH) mass function in the range of $(10^4, 10^7) M_\odot$,
\be 
\frac{dN_\bullet}{d\log M_\bullet} = 0.01\left(\frac{M_\bullet}{3\times 10^6 M_\bullet}\right)^{-0.3}\ {\rm Mpc}^{-3}\ ,
\ee 
following Refs.~\cite{Babak2017,Pan2021b}.  
The differential EMRI rate in the dry channel is written as 
\be 
\frac{d^2\mathcal{ R}}{dM_\bullet dz} = \frac{1}{1+z}\frac{dN_\bullet}{dM_\bullet}\frac{dV_{\rm c}(z)}{dz} C_{\rm cusp}(M_\bullet, z) \Gamma (M_\bullet, N_{\rm p})\ ,
\ee 
where $V_{\rm c}(z)$ is the  comoving volume of the universe up to redshift $z$, $C_{\rm cusp}(M_\bullet, z) $ the fraction of MBHs living in stellar
cusps (see \cite{Babak2017} for calculation details) and $\Gamma(M_\bullet, N_{\rm p})$ is the generic EMRI formation rate in the dry channel, 
with $N_{\rm p} (\approx 10)$ the average number of plunges per EMRI. In the formula above, dry EMRIs and Hills EMRIs are not distinguished, due to the large uncertainty 
in  the rate of Hills EMRIs. As a simple parametrization, we write the differential rates in the two channels as 
\be 
\begin{aligned}
    \frac{d^2\mathcal{ R_{\rm dry}}}{dM_\bullet dz} &= f_{\rm dry} \frac{d^2\mathcal{R}}{dM_\bullet dz}\ ,\\
\frac{d^2\mathcal{ R_{\rm Hills}}}{dM_\bullet dz} &= (1- f_{\rm dry}) \frac{d^2\mathcal{R}}{dM_\bullet dz}\ .
\end{aligned}
\ee 

For calculating the detectable rate, we fix the dimensionless spin of the MBH as $a=0.99$, the secondary BH mass as $m=10M_\odot$. The MBH mass $M_\bullet$ is randomly sampled according to the differential EMRI rates, EMRI eccentricity $e_0$ at coalescence is sampled from a uniform distribution in $\left[0,0.2\right]$ and $\cos{\iota_0}$ is  
sampled from a uniform distribution in $\left[-1,1\right]$. 
Instead of using the initial semi-latus rectum, we choose the coalescence time $T$ which is randomly sampled from $\left[0,2\right]\text{yr}$, and integrate backwards from Kerr last stable orbit (LSO) with duration $T$. For extrinsic parameters, $\hat{N}$ and $\hat{A}$ are isotropically distributed on the celestial sphere, $\Phi$ is uniformly distributed in $\left[0,2\pi\right]$. After sampling 1300 ``raw'' events, we select those that are detectable by LISA or Taiji ($\rho \geq 20$), and there are eventually $N_{\rm det}=121$ EMRIs left.
When sampling the 2nd spin $s$ from Eq.~(\ref{total_spin}), we take the ``true'' values of $\mathbf{\Lambda}$ as $\mathbf{\Lambda}_0 = \left\{0.1,-4,0.4,0.1,0.5\right\}$. Finally, 57 EMRIs are randomly selected as dry ones from above 121 events, and the other 64 EMRIs are left to be Hills ones.

% Then we plug $\boldsymbol{\lambda}$ of each EMRI event into our waveform module to generate waveforms and calculate the corresponding Fisher information matrix in Eq.\ref{eq:detectlikelihood}.
% And the best-fit population parameters $\mathbf{\Lambda}_\text{bf}$ should be those maximizing the likelihood $\mathcal{L}\left(\mathbf{\Lambda} \mid \{\mathbf{d_i}\}\right) \propto \mathcal{L}\left(\{\mathbf{d}_i\} \mid \mathbf{\Lambda}\right)$.

\begin{figure}
    \centering
    \includegraphics[width=1.\linewidth]{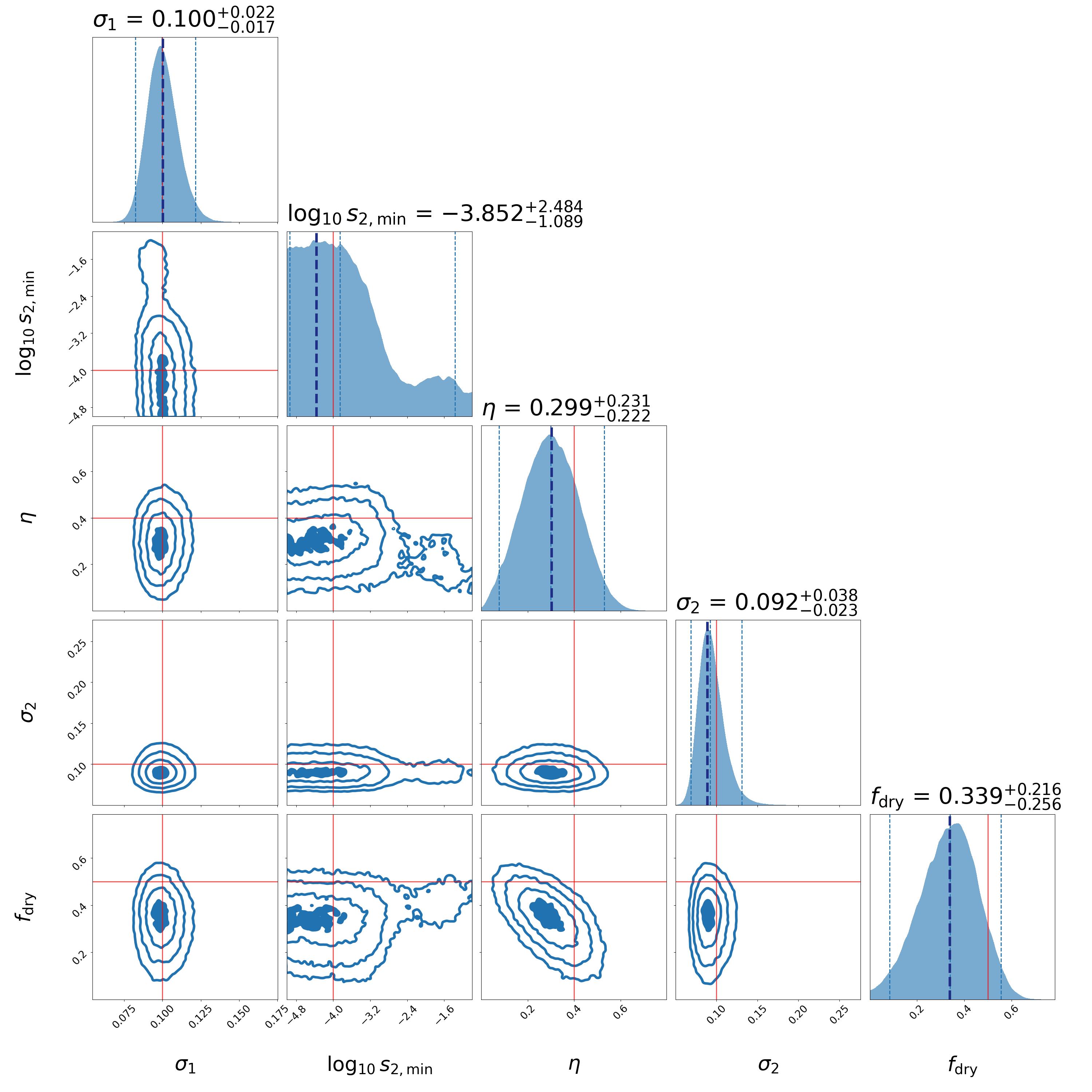}
    \caption{Posterior distribution of the population parameters $\mathbf{\Lambda}$, 
    where the red solid lines mark the true value $\mathbf{\Lambda}_0$, the dark dotted lines mark the best-fit,
and the light dotted lines mark the $95\%$ credible interval.} 
    \label{fig:dry_hill_mix_1}
\end{figure}

With the $N_{\rm det}$ EMRIs that are detectable,  we perform a population inference with the likelihood defined in Eq.~(\ref{eq:PopLikelihood}) 
using the dynamic nested sampling algorithm \cite{dynesty}.
The posterior distribution of the population parameters $\mathbf{\Lambda}$ is shown in 
Fig.~\ref{fig:dry_hill_mix_1}.
% Firstly, it's obvious that all $\Lambda^i_0$ are within 95\% credible interval in our simulation, and the most interesting parameter $f_\text{dry}$ is inferred accurately as 
% \be
% \mid f_\text{dry,0}-f_\text{dry,bf}\mid/f_\text{dry,0} \approx 1.2\% ,
% \ee
% but not constrained very well as the 95\% credible interval is somewhat broad, 
% \be
% \Delta_{95\%} f_\text{dry}/f_\text{dry,0} \approx 64.3\% . 
% \ee
where $\sigma_1$ is inferred better than $\sigma_2$, as result of the fact that our injection events contains more information about $P_1(s)$ than $P_\text{sync}(s)$ from Eq.~(\ref{total_spin}), as $\eta_{P_1}=(1+f_\text{dry,0})/2=0.75$ and $\eta_{P_\text{sync}}=(1-f_\text{dry,0})(1-\eta_0)/2=0.15$. 
For $s_\text{2,min}$, it represents a minimal limit of second born BH spins in non-synchronization Hills mechanism, and has little effect on the injection as it's considered to be extremely low value, $10^{-4}$, which should seldom be reached when sampling, so $\log_{10} s_\text{2,min}$ is constrained poorly in our inference.

\begin{figure}
    \centering
    \includegraphics[width=1.\linewidth]{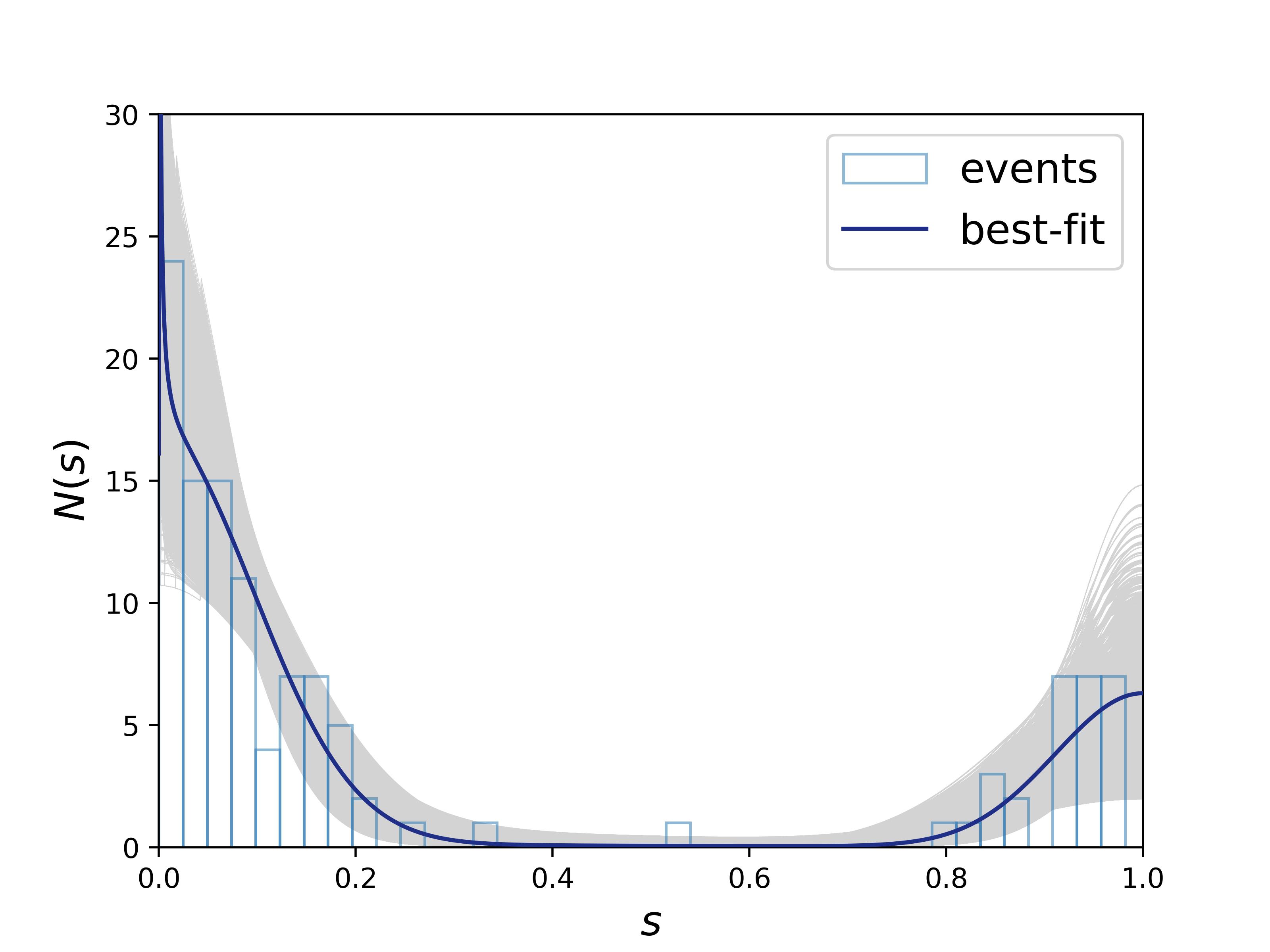}
    \caption{Distributions of secondary spins $s$ of dry and Hills EMRIs: injected (histogram) and recovered (best-fit in black and $95\%$ credible interval in gray).}
    %where the high-spin peak is a signature of second born BHs in Hills EMRIs.}
    \label{fig:events_fit}
\end{figure}

In Fig.~\ref{fig:events_fit}, we compare the injected secondary spins (light blue histogram) and the recovered distribution obtained from the population inference. 
The bundle of light gray curves represent $P(s; \mathbf{\Lambda})$ with inferred population parameters $\mathbf{\Lambda}$ within the 95\% credible interval. The high-spin peak should be a unique signature of second-born BHs in Hills EMRIs, while both dry and Hills EMRIs will contribute to the low-spin peak.

\section{Conclusion and Discussion}\label{section:conclusion}
In this paper, we constructed a kludge EMRI waveform model that includes the effect of the secondary spin by numerically integrating 
the MPD equations (\ref{MPD1}-\ref{MPD3}) and PN radiation reactions (\ref{RR}). With this waveform template, we forecast the LISA measurement precision of secondary spins of EMRIs using the Fisher information matrix. As a result, we found that the secondary spin in general is poorly constrained for low-eccentricity and nearly-equatorial EMRIs, 
but is much better constrained for generic eccentric and/or inclined EMRIs (see Fig.~\ref{fig:heatmap}), 
due to stronger coupling between the secondary spin and EMRI orbital dynamics in this case. 
\textcolor{red}{
Note that we certainly won't expect this rudimentary template to achieve the requisite precision for future waveform-matched filtering, and the insufficiently verified use of Fisher-Matrix analysis in this paper may also introduce uncertainties to the quantitative results.
}

With future EMRI detections, 
the stellar mass BH formation history and the EMRI formation process 
encoded by the secondary spin can be extracted and used as a probe to stellar evolution and stellar dynamics in nuclear clusters.
As a simple example of such astrophysical applications, we did a test on whether dry EMRIs and Hills EMRIs can be distinguished via the secondary spin. Previous studies show that dry and Hills EMRIs are similar in their orbital parameters on the population level, high eccentricities and isotropic inclinations. In this work, we pointed out that the secondary spins of dry and Hill EMRIs are expected to be 
quite different:
the sBH from the dry channel keeps the natal low spin as born in the collapse of a massive star  \cite{LIGOScientific:2018jsj,Roulet:2018jbe,Fuller2019,KAGRA:2021duu}, 
while the sBH from the Hills channel might be spun up during the He-core stage by its companion \cite{Zaldarriaga:2017qkw,Bavera:2020inc,Mandel:2020lhv}. Therefore, high secondary spins will be a clear signature of Hills EMRIs.
As a concrete example, we parameterized the secondary spin distributions of dry and Hills EMRIs as in Eq.~(\ref{dry_spin}) and Eq.~(\ref{Hills_spin}), respectively. 
Via an injection and inference experiment, we found that several population parameters, 
including the secondary spin uncertainty $\sigma_1$ and the branch ratio $f_{\rm dry}$ of the dry channel,
are expected to be well measured from a reasonable number of EMRI detections ($N_{\rm det}=121$). 
Of course, this conclusion depends on the validity of Eqs.~(\ref{dry_spin},\ref{Hills_spin}).
On the other hand, if these simple parameterizations are not good approximations 
to the true distributions, 
the robust detection of the high-spin peak will still be a clear signature of Hills EMRIs,
though  we may not trust the exact interpretation of individual population parameters in this case.

Wet EMRIs are expected to be of low eccentricity in the LISA sensitivity band due to the efficient eccentricity damping by density waves as migrating along the accretion disk.
If AGN accretion is coherent, wet EMRIs that are formed in the AGN disk are preferentially on the equator. In this case, the secondary spins are poorly constrained (see Fig.~\ref{fig:heatmap}). If AGN accretion is chaotic with gas feeding from a random direction in each accretion episode, 
wet EMRIs that are brought in by the outer disk are expected to be misaligned with the inner equatorial disk \cite{Lyu:2024gnk}. 
In this case, the secondary spins could be constrained (Fig.~\ref{fig:heatmap}) and used as probe to the accretion history of EMRIs embedded in the AGN disk \cite{Pan2021,Li:2022dqh,Chen:2023lxk}.

\begin{acknowledgments}
We thank Ying Qin for helpful discussions on the binary star evolution.
This work is supported by the National Key R\&D Program of China (Grant No. 2021YFC2203002), NSFC (National Natural Science Foundation of China) No. 12473075, No. 12173071 and No. 11773059. This work made use of the High Performance Computing Resource in the Core Facility for Advanced Research Computing at Shanghai Astronomical Observatory.
\end{acknowledgments}
\appendix

\bibliography{main}
\end{document}